\documentclass[12pt]{article}
\usepackage{amssymb,amsmath,amsfonts,eurosym,geometry,ulem,graphicx,caption,color,setspace,sectsty,comment,footmisc,caption,natbib,pdflscape,array,hyperref}
\usepackage{amsthm}
\newtheorem{assumption}{Assumption}

\usepackage{float}        
\usepackage{threeparttable} 
\usepackage{multirow}     
\usepackage{subcaption}
\usepackage{authblk} 
\usepackage{graphicx}
\usepackage{caption}

\usepackage{alphalph} 

\hypersetup{
	colorlinks=true,      
	linkcolor=blue,       
	citecolor=blue,       
	filecolor=blue,       
	urlcolor=blue         
}

\newtheorem{theorem}{Theorem}

\newcolumntype{L}[1]{>{\raggedright\let\newline\\arraybackslash\hspace{0pt}}m{#1}}
\newcolumntype{C}[1]{>{\centering\let\newline\\arraybackslash\hspace{0pt}}m{#1}}
\newcolumntype{R}[1]{>{\raggedleft\let\newline\\arraybackslash\hspace{0pt}}m{#1}}

\title{Wild Bootstrap Inference for Linear Regressions with Many Covariates}
\author{Wenze Li\thanks{Division of Economics, School of Social Sciences, Nanyang Technological University. Email: \texttt{wenze001@e.ntu.edu.sg}. \\

\quad I am sincerely grateful to Wenjie Wang (Nanyang Technological University) for his generous guidance and insightful comments, which have substantially strengthened the arguments and analysis presented in this paper.}}

\date{June 2025}

\begin{document}

\maketitle

\maketitle

\begin{abstract}

We propose a simple modification to the wild bootstrap procedure and establish its asymptotic validity for linear regression models with many covariates and heteroskedastic errors. Monte Carlo simulations show that the modified wild bootstrap has excellent finite sample performance compared with alternative methods that are based on standard normal critical values, especially when the sample size is small and/or the number of controls is of the same order of magnitude as the sample size.  
\\

     \bigskip
   

\end{abstract}

\section{Introduction}

An important econometric method for estimating the causal effect of a certain variable is based on the linear regression model with the assumption that the variable of interest is exogenous after controlling for a sufficiently large set of covariates (e.g., panel data models with fixed effects or partial linear models). 
However, inference with many covariates can be challenging under (conditional) heteroskedasticity.   
\cite{cattaneo2018alternative} provide a unified framework for nonstandard distributional approximation, which accommodates a variety of asymptotics studied in the literature, including the small bandwidth, many instruments, and many covariates asymptotics.
Furthermore, as shown by \cite{cattaneo2018inference}, under the asymptotic framework where the number of covariates, $q_n$, grows with the sample size, $n$, the consistency of the traditional heteroskedasticity-robust variance estimators 
would require $q_n/n \rightarrow 0$. 
In their seminal studies, \cite{cattaneo2018inference} and \cite{jochmans2022heteroscedasticity} propose new variance estimators that remain consistent even when $q_n/n \nrightarrow 0$,
and provide asymptotically valid inferences that are based on normal critical values.

However, it is well known that asymptotic normal approximations may have finite sample distortions, especially when the number of observations is small. 
Bootstrap is found in the econometric literature to provide an improvement over asymptotic approximations in many situations; e.g., see \cite{Hall-Horowitz(1996)}, \cite{Horowitz(2001)}, 
\cite{Davidson-Mackinnon(2010)}, \cite{Djogbenou-Mackinnon-Nielsen(2019)}, and the references therein.    
However, it may fail in high-dimensional settings; e.g., see \cite{el2018can} and the references therein. 
For linear regressions with heteroskedastic errors, \cite{mammen1993bootstrap} established that the validity of nonparametric bootstrap and wild bootstrap requires $q^{1+\delta}_n/n \rightarrow 0, \delta>0$. Similarly, for linear instrumental variable (IV) models with many IVs and homoskedastic errors, \cite{Wang-Kaffo(2016)} show bootstrap inconsistency for the standard residual-based bootstrap when the number of IVs is of the same order of magnitude as the sample size $n$, and propose a valid alternative bootstrap procedure. However, their procedure cannot be directly extended to the case with both many IVs and heteroskedasticity, even with a fixed number of controls.  

To our knowledge, there is no existing proven-valid bootstrap method for linear regression models under both $q_n/n \nrightarrow 0$ and heteroskedasticity. Simulations in \cite{cattaneo2018inference} also suggest that the nonparametric bootstrap seems to fail in their settings (e.g., see Remark 2 in their paper). 
In this paper, following the construction of the (approximate) cross-fit variance estimator in \cite{jochmans2022heteroscedasticity}, we propose a simple modification to the standard wild bootstrap and show its asymptotic validity when $q_n/n \nrightarrow 0$.   
Monte Carlo experiments illustrate good finite sample performance of our procedure. 

\section{Model and Setup}\label{sec: setup}

Consider the linear regression model 
\begin{align}\label{eq: model}
y_{i} = x_{i} \beta + w'_{i} \gamma + u_{i}, \; i=1, ..., n, 
\end{align}
where $y_{i}$ is a scalar outcome variable, $x_{i}$ is a scalar explanatory variable such as a certain treatment, $w_{i}$ is a $q_n \times 1$ vector of covariates,  and $u_{i}$ is an error term. 
We allow $q_n$ to be a non-negligible fraction of the sample size $n$. 
The ordinary least squares (OLS) estimator of $\beta$ is then defined as
\begin{align*}
\hat{\beta}_n = \left( \sum_{i=1}^n \hat{v}^2_{i} \right)^{-1} \left( \sum_{i=1}^n \hat{v}_{i} y_{i} \right), 
\end{align*}
where $\hat{v}_{i} = \sum_{j=1}^n (M_n)_{ij} x_{j}$,
$(M_n)_{ij} = \{ i=j \} - w_i' (\sum_{k=1}^n w_kw_k')^{-1} w_j$, 
and $\{ \cdot \}$ denotes the indicator function. 

Recently, \cite{cattaneo2018inference} propose a new heteroskedasticity-robust variance estimator, which is consistent when $\limsup_{n} q_n/n < 1/2$. 
Furthermore, \cite{jochmans2022heteroscedasticity} proposes an alternative variance estimator, which remains consistent as long as $\limsup_{n} q_n/n < 1$, and its 
corresponding $t$-ratio can be defined as 
$t_n = (\hat{\beta}_n - \beta_0)/\sqrt{\acute{\Omega}_n}$, where 
\begin{align}\label{eq: joch}
\acute{\Omega}_n = \left(\sum_{i=1}^n \hat{v}_i^2\right)^{-1} \left( \sum_{i=1}^n \hat{v}_i^2 (y_i \acute{u}_i) \right) \left(\sum_{i=1}^n \hat{v}_i^2\right)^{-1},
\end{align}
with $\acute{u}_i = \hat{u}_i/(M_n)_{ii}$ and $\hat{u}_i = \sum_{j=1}^n (M_n)_{ij} (y_j - x_j \hat{\beta}_n)$.
Then, $t$-tests based on normal critical values have correct size in large samples. 
However, simulations suggest that asymptotic normal approximation may not have satisfactory finite sample performance (Section \ref{sec: simu}), especially when the sample size is small and/or $q_n$ equals a large fraction of $n$. 
In Section \ref{sec: score boot}, we propose a modified wild bootstrap procedure, which follows the (approximate) cross-fit variance estimator of \cite{jochmans2022heteroscedasticity}.   

\section{Modified Wild Bootstrap Procedure}\label{sec: score boot}

Our procedure is defined as follows: 

\textbf{Step 1:} Given the null hypothesis $H_0: \beta=\beta_0$, 
generate the residuals $\{ \tilde{u}_i(\beta_0) \}_{i=1}^n$, where 
$\tilde{u}_i(\beta_0) = y_i - x_i \beta_0 - w_i' \tilde{\gamma}_n,$ 
and $\tilde{\gamma}_n$ is the null-restricted OLS estimator of $\gamma$.

\textbf{Step 2:} Generate $\{ u_i^* \}_{i=1}^n$, where $u_i^* = a_n(\beta_0) \omega^*_i \tilde{u}_i(\beta_0)$, 
$\{ \omega^*_i \}_{i=1}^n $ is an i.i.d. sample of random weights that are independent of the data with zero mean and unit variance while 
\begin{align}
a_n(\beta_0) = \sqrt{\max\{ \acute{\Sigma}_n(\beta_0), 1/n\} / \hat{\Sigma}_n(\beta_0)}
\end{align}
is an adjustment factor, 
which takes into account the high dimensionality of the covariates. 
Specifically, $\acute{\Sigma}_n(\beta_0) = \sum_{i=1}^n \hat{v}_i^2 y_i \acute{u}_i(\beta_0)$, 
$\hat{\Sigma}_n(\beta_0) = \sum_{i=1}^n \hat{v}_i^2 \tilde{u}^2_i(\beta_0)$,
$\tilde{u}_i(\beta_0)$ is defined in Step 1, and $\acute{u}_i(\beta_0) = \tilde{u}_i(\beta_0)/(M_n)_{ii}$. 
Notice that $\acute{\Sigma}_n(\beta_0)$ is similar to the ``meat" part of $\acute{\Omega}_n$ in (\ref{eq: joch}), but here we impose $H_0$, as we generated the bootstrap samples with the null imposed in Step 1. Additionally, $max\{ \cdot, 1/n \}$ guards against the case that $\acute{\Sigma}_n(\beta_0)$ could take a non-positive value in finite samples.\footnote{This may also occur for the variance estimators of \cite{cattaneo2018inference} and \cite{jochmans2022heteroscedasticity}.}

\textbf{Step 3:} Generate $y^*_i = x_i \beta_0 + w_i' \tilde{\gamma}_n + u_i^*, i=1, ..., n$. 
Then, compute the bootstrap OLS estimator $\hat{\beta}^*_n = \left( \sum_{i=1}^n \hat{v}^2_{i} \right)^{-1} \left( \sum_{i=1}^n \hat{v}_{i} y^*_{i} \right)$,
and the bootstrap residual $\hat{u}_i^* = \sum_{j=1}^n (M_n)_{ij}(y^*_j - x_j \hat{\beta}^*)$.

\textbf{Step 4:} Compute $t_n^* = (\hat{\beta}^*_n - \beta_0)/\sqrt{\acute{\Omega}^*_n}$, 
where $\acute{\Omega}^*_n = (\sum_{i=1}^n \hat{v}_i^2)^{-1} (\sum_{i=1}^n \hat{v}_i^2 (y^*_i \acute{u}^*_i)) (\sum_{i=1}^n \hat{v}_i^2)^{-1}$, with $\acute{u}_i^* = \hat{u}_i^*/(M_n)_{ii}$, i.e., $\acute{\Omega}^*_n$ has the same formula as $\acute{\Omega}_n$ but uses bootstrap samples. 
 
\textbf{Step 5:} Repeat Steps 2-4 B times, and compute the bootstrap $p$-value
$p_n^* = B^{-1}\sum_{b=1}^B 1\{ |t_n| > |t_n^{*(b)}| \}$.
We reject $H_0$ if $p_n^*$ is smaller than the nominal level $\alpha$.

Several comments are in order. 

\textbf{Comment 1.} In Step 1, we impose null when computing the residuals, which is advocated by \cite{Cameron(2008)}, \cite{Davidson-Flachaire(2008)}, \cite{Roodman-Nielsen-MacKinnon-Webb(2019)}, among others. 
As a consequence, we also impose null for the adjustment factor $a_n(\beta_0)$ in Step 2. When $q_n/n \rightarrow 0$, $\acute{\Sigma}_n(\beta_0)$ and $\hat{\Sigma}_n(\beta_0)$ are asymptotically equivalent so that our procedure reduces to the standard (null-imposed) wild bootstrap. 

\textbf{Comment 2.} Instead of the above percentile-$t$ type procedure, one may consider a percentile type procedure, whose bootstrap $p$-value equals
$B^{-1}\sum_{b=1}^B 1\{ | \sqrt{n}(\hat{\beta}_n - \beta_0)| > | \sqrt{n}(\hat{\beta}^{*(b)}_n - \beta_0)|\}$.
However, in line with the recommendation in the literature (e.g., see the papers cited above), we find in simulations that the percentile-$t$ has better behaviours and thus also recommend using it.  

\textbf{Comment 3.} Here we focus on the case where $x_{i}$ is a scalar variable. It is possible to extend the analysis to the case where $x_i$ is a (fixed) $d_x$-dimensional vector by considering a modification to the score bootstrap (e.g., \cite{Kline-Santos(2012)}). 
Specifically, the procedure for testing $H_\lambda: c'\beta = \lambda$, where $c \in R^{d_x}$ and $\lambda \in R$, can be defined as follows: 

(1) Obtain the null-restricted OLS estimator $\tilde{\beta}_n$ and $\tilde{\gamma}_n$, and employ it to the score contributions $\{ S_i(\tilde{\beta}_n) \}_{i=1}^n$, where $S_i(\tilde{\beta}_n) = \hat{v}_i\tilde{u}_i(\tilde{\beta}_n)$ and $\tilde{u}_i(\tilde{\beta}_n) = y_i - x'_i\tilde{\beta}_n - w_i'\tilde{\gamma}_n$; 

(2) Compute the perturbed score contributions $\{ S_i^*(\tilde{\beta}_n) \}_{i=1}^n$, 
where $S_i^*(\tilde{\beta}_n) = \omega^*_i \hat{v}_i \tilde{u}_i(\tilde{\beta}_n)$; 

(3) Compute the bootstrap statistic
$T_n^* = c' \left( n^{-1} \sum_{i=1}^n \hat{v}_i\hat{v}_i' \right)^{-1} \hat{A}_n(\tilde{\beta}_n) 
( n^{-1/2} \sum_{i=1}^n S_i^*(\tilde{\beta}_n))$,
where $\hat{A}_n(\tilde{\beta}_n) = \acute{\Sigma}^{1/2}_n(\tilde{\beta}_n)\hat{\Sigma}^{-1/2}_n(\tilde{\beta}_n)$ is an adjustment matrix with $\acute{\Sigma}_n(\tilde{\beta}_n) = \sum_{i=1}^n \hat{v}_i \hat{v}_i' y_i \acute{u}_i(\tilde{\beta}_n)$ and 
$\hat{\Sigma}_n(\tilde{\beta}_n) = \sum_{i=1}^n \hat{v}_i \hat{v}_i' \tilde{u}^2_i(\tilde{\beta}_n)$; 

(4) Use the distribution of $T_n^*$ conditional on the data as an estimate of the null distribution of 
$T_n = \sqrt{n} ( c'\hat{\beta}_n - \lambda )$.

When $x_i$ is a scalar, this modified score bootstrap is equivalent to the percentile-type procedure in Comment 2. 
We recommend the current procedure in Steps 1-5 as it has better finite sample performance and might also be more user-friendly to practitioners (e.g., simply need to add an adjustment factor to the residuals for the standard wild bootstrap).

\section{Asymptotic Theory}\label{sec:theory}

Below we introduce some regularity conditions needed to establish the bootstrap validity. Assumptions \ref{ass: 1}-\ref{ass: 4} follow closely those in \cite{cattaneo2018inference} and \cite{jochmans2022heteroscedasticity}.
Assumption \ref{ass: 5} contains conditions for the bootstrap procedure.
Appendix \ref{sec:diss_assumption} gives further discussions on these assumptions.

Let $\mathcal{W}_n$ denote a collection of random variables such that $E[w_{i} | \mathcal{W}_n] = w_{i}$.
Let $\epsilon_{i} = u_{i} - e_{i}$, where $e_{i} = E [u_{i} | X_n, \mathcal{W}_n]$, and $V_{i} = x_{i} - E [x_{i} | \mathcal{W}_n]$. 
Let $\sigma_{i}^2 = E [\epsilon_{i}^2 | X_n, \mathcal{W}_n]$, 
and $\tilde{V}_{i} = \sum_{j=1}^n (M_n)_{ij} V_{j}$. 

\begin{assumption}\label{ass: 1} 
The errors $\epsilon_i$ are uncorrelated across $i$ conditional on $X_n$ and $\mathcal{W}_n$, and the collections $\{ \epsilon_i, V_i: i \in N_g\}$ are independent across $g$ conditional on $\mathcal{W}_n$,  where $N_1, ..., N_G$ represents a partition of $\{1, ..., n\}$ into $G$ sets such that $\max_g|N_g| = O(1)$.
\end{assumption}

\begin{assumption}\label{ass: 2} 
With probability approaching one, $\sum_{i=1}^n w_{i}w'_{i}$ has full rank, 
\begin{align*}
\max_{i} \left( E[\epsilon_{i}^4 | X_n,  \mathcal{W}_n] + \sigma_{i}^{-2} + E[V_{i}^4| \mathcal{W}_n ]\right) +  
\left(\lambda_{min} \left(n^{-1} \sum_{i=1}^n E[ \tilde{V}^2_{i} | \mathcal{W}_n ] \right)\right)^{-1} = O_p(1), 
\end{align*}
and $\limsup_{n} q_n /n <1$, where $\lambda_{min}(\cdot)$ denotes the minimum eigenvalue of its argument. 
\end{assumption}

\begin{assumption}\label{ass: 3}
$\chi_n = O(1), \eta_n + n(\eta_n - \rho_n) + n \chi_n \eta_n = o(1), $
and $\max_{i} |\hat{v}_{i} | /\sqrt{n} = o_p(1)$, 
where $\eta_n = n^{-1} \sum_{i=1}^n E[e_{i}^2]$, $\rho_n = n^{-1} \sum_{i=1}^n E[ E[e_{i} | \mathcal{W}_n]^2]$, $\chi_n = n^{-1} \sum_{i=1}^n E[ Q_{i}^2]$, $Q_{i} = E[v_{i} | \mathcal{W}_n]$, 
and $v_i = x_i - (\sum_{j=1}^n E[x_j w_j'])(\sum_{j=1}^n E[w_jw_j'])^{-1}w_i$.  
\end{assumption}

\begin{assumption}\label{ass: 4}
$n \eta_n = O(1), $ $P\left[ \min_i (M_n)_{ii} > 0 \right] \rightarrow 1$, 
$( \min_i (M_n)_{ii} )^{-1} = O_p(1)$,  $n^{-1} \sum_{i=1}^n \tilde{Q}^4_{i} = O_p(1)$, 
and $\max_i |\mu_{i} | /\sqrt{n} = o_p(1)$, 
where $\mu_{i} = E[y_{i} | X_n, \mathcal{W}_n]$ and $\tilde{Q}_{i} = \sum_{j=1}^n (M_n)_{i,j} Q_{j}$.
\end{assumption}

\begin{assumption}\label{ass: 5}
$\{ \omega^*_i \}_{i=1}^n$ is an i.i.d. sample of random weights independent of the data, 
and it satisfies $E[\omega^*_i]=0$, $E[\omega_i^{*2}]=1$, and $E[\omega_i^{*4}] < \infty$.
Furthermore, $\max_i |\hat{\mu}_{i} (\beta_0) | /\sqrt{n} = o_p(1)$ under the null,  where $\hat{\mu}_i(\beta_0) = x_i \beta_0 + w_i'\tilde{\gamma}_n$.
\end{assumption}

\begin{theorem}\label{theom: 1}
Let $F_n$ and $F_n^*$ be the cdfs of $t_n$ and of $t_n^*$ conditional on the sample,  respectively.
Suppose that Assumptions \ref{ass: 1}-\ref{ass: 5} hold. If $H_0$ is true, then
\begin{align*}
\sup_{c \in R} |F_n(c) - F_n^*(c)| = o_p(1). 
\end{align*}
\end{theorem}

Theorem \ref{theom: 1} shows the asymptotic validity of the modified wild bootstrap under many covariates and heteroskedasticity.  For the proof of the bootstrap central limit theorem (CLT),  we apply the conditional multiplier CLT (e.g., Chapter 2.9 of \cite{van1996weak}).  

\section{Monte Carlo Simulations}\label{sec: simu}

The model is similar to that considered in \cite{cattaneo2018inference} and
\cite{jochmans2022heteroscedasticity}: 
\begin{align*}
y_{i} = x_{i}\beta + w'_{i}\gamma + \epsilon_{i}, 
\end{align*}
where $x_{i} \sim i.i.d. N(0,1)$, $w_{i}$ contains a constant term and a collection of $q_n-1$ zero/one dummy variables, and $\epsilon_{i} \sim i.i.d. N(0,1)$.
The dummy variables are drawn independently with success probability $\pi$ and $\gamma=0$. 
The sample size is fixed to $n=100$, and $q_n/n \in \{0.1, 0.2, 0.3, 0.4, 0.5, 0.6, 0.7, 0.8, 0.9\}$. The number of Monte Carlo replications is set to $10,000$ and the number of bootstrap replications is set to $199$ throughout the simulations. 
Following \cite{jochmans2022heteroscedasticity}, we consider three designs that vary in $\beta$ and $\pi$: $\beta=1$ and $\pi=0.02$ (Design A), $\beta=1$ and $\pi=0.01$ (Design B), 
and $\beta=2$ and $\pi=0.02$ (Design C). 
The results for Design A are presented in Table \ref{tab:1}.
The empirical null rejection frequencies of the two-sided $t$-test are given for the Eicker-White variance estimator (``HC0"), the variance estimator of \cite{cattaneo2018inference} (``HCK"), 
the variance estimator of \cite{jochmans2022heteroscedasticity} (``HCA"), 
and the proposed wild bootstrap with Gaussian and Rademacher random weights (``Wild-G" and ``Wild-R"), respectively. 

We highlight several findings below. The tests based on ``HC0",  ``HCK",  and ``HCA" all tend to over-reject as $q_n$ increases, with ``HCA" having the smallest distortions among the three (e.g., when $q_n/n=0.9$, their null rejection frequencies
are $58.1\%$, $58.1\%$, and $17.2\%$, respectively). 
By contrast, both bootstraps control size across various values of $q_n$.
Also, both become more conservative when $q_n$ increases, and  
the one with Gaussian weights seems to be slightly less conservative than that with Rademacher weights.
The results for Design B and Design C are reported in Table \ref{tab:2} and Table \ref{tab:3}, respectively. 
The overall patterns are very similar to those observed in Table \ref{tab:1}.
In Section \ref{sec:further simu} of the Supplementary Appendix, we present further simulation results for a panel data model.  


\begin{table}[H]
\begin{center}
\begin{tabular}{llllllllll}
$q_n/n$ & 0.1 & 0.2 & 0.3 & 0.4 & 0.5 & 0.6 & 0.7 & 0.8 & 0.9 \\ 
\hline
\hline
HC0    & 0.071 & 0.097 & 0.116 & 0.150 & 0.186 & 0.243 & 0.316 & 0.407 & 0.581 \\
HCK    & 0.059 & 0.069 & 0.071 & 0.083 & 0.095 & 0.116 & 0.165 & 0.228 & 0.581 \\
HCA    & 0.067 & 0.075 & 0.076 & 0.084 & 0.084 & 0.087 & 0.100 & 0.121 & 0.172 \\
Wild-G & 0.051 & 0.046 & 0.047 & 0.048 & 0.045 & 0.044 & 0.044 & 0.040 & 0.035 \\
Wild-R & 0.038 & 0.044 & 0.036 & 0.041 & 0.040 & 0.039 & 0.041 & 0.039 & 0.033 \\
\hline
\end{tabular}
\caption{Null rejection frequencies for Design A}
\label{tab:1}
\end{center}
\footnotesize{Note: ``HC0", ``HCK", and ``HCA" denote the variance estimators of Eicker-White, \cite{cattaneo2018inference}, and \cite{jochmans2022heteroscedasticity}, respectively, 
while ``Wild-G" and ``Wild-R" denote the proposed wild bootstrap with Gaussian and Rademacher weights, respectively.}
\end{table}

\begin{table}[H]
\begin{center}
\begin{tabular}{llllllllll}
$q_n/n$ & 0.1 & 0.2 & 0.3 & 0.4 & 0.5 & 0.6 & 0.7 & 0.8 & 0.9 \\ 
\hline
\hline
HC0    & 0.075 & 0.097 & 0.113 & 0.152 & 0.189 & 0.242 & 0.314 & 0.410 & 0.574 \\
HCK    & 0.062 & 0.069 & 0.070 & 0.081 & 0.092 & 0.117 & 0.168 & 0.238 & 0.574 \\
HCA    & 0.072 & 0.075 & 0.075 & 0.082 & 0.084 & 0.090 & 0.107 & 0.127 & 0.176 \\
Wild-G & 0.050 & 0.050 & 0.045 & 0.048 & 0.044 & 0.045 & 0.042 & 0.040 & 0.037 \\
Wild-R & 0.037 & 0.041 & 0.036 & 0.041 & 0.036 & 0.040 & 0.037 & 0.036 & 0.033 \\
\hline
\end{tabular}
\caption{Null rejection frequencies for Design B}
\label{tab:2}
\end{center}
\end{table}

\begin{table}[H]
\begin{center}
\begin{tabular}{llllllllll}
$q_n/n$ & 0.1 & 0.2 & 0.3 & 0.4 & 0.5 & 0.6 & 0.7 & 0.8 & 0.9 \\ 
\hline
\hline
HC0    & 0.073 & 0.095 & 0.122 & 0.141 & 0.195 & 0.240 & 0.310 & 0.413 & 0.583 \\
HCK    & 0.063 & 0.067 & 0.072 & 0.075 & 0.095 & 0.120 & 0.161 & 0.233 & 0.583 \\
HCA    & 0.097 & 0.105 & 0.114 & 0.104 & 0.120 & 0.124 & 0.136 & 0.151 & 0.175 \\
Wild-G & 0.044 & 0.041 & 0.043 & 0.036 & 0.042 & 0.042 & 0.041 & 0.040 & 0.036 \\
Wild-R & 0.037 & 0.034 & 0.034 & 0.031 & 0.034 & 0.037 & 0.041 & 0.036 & 0.031 \\
\hline
\end{tabular}
\caption{Null rejection frequencies for Design C}
\label{tab:3}
\end{center}
\end{table}

\section{Conclusion}

In this paper, we propose a simple modification to the standard wild bootstrap procedure for linear regression models with many covariates and heteroskedastic errors. We establish its asymptotic validity when the number of covariates is of the same order of magnitude as the sample size. 
Our construction of the adjustment factor in the bootstrap procedure follows that of the (approximate) cross-fit variance estimator proposed by \cite{jochmans2022heteroscedasticity}. 
Monte Carlo simulations show that the modified wild bootstrap has excellent finite sample performance compared with alternative methods that are based on standard normal critical values, especially when the sample size is small and/or the number of covariates is of the same order of magnitude as the sample size.  
For potential directions for future research, we note that there is a growing literature on robust inference under many (weak) instruments and non-homoskedastic errors, possibly also with many controls.\footnote{E.g., see \cite{evdokimov2018inference}, \cite{crudu2021}, \cite{MS22}, \cite{matsushita2024jackknife}, \cite{LWZ(2023)}, \cite{DKM24},  \cite{boot-ligtenberg(2023)}, \cite{N23}, \cite{boot2024}, \cite{lim2024dimension}, and \cite{yap2024inference}, among others.}
Due to the complexity of data structure, it is possible that asymptotic normal approximations may have less than satisfactory performance in this setting as well. 
On the other hand, it is found that when implemented appropriately, bootstrap approaches may substantially improve the inference accuracy for IV models, including the cases where IVs may be rather weak.\footnote{E.g., see \cite{Moreira-Porter-Suarez(2009)}, 
\cite{Davidson-Mackinnon(2008), Davidson-Mackinnon(2010), Davidson-Mackinnon(2014b)}, \cite{wang2015bootstrap},
\cite{Wang-Kaffo(2016)}, \cite{Kaffo-Wang(2017)}, \cite{Wang-Doko(2018)}, \cite{Finlay-Magnusson(2019)}, \cite{mackinnon2023fast}, and \cite{Wang-Zhang2024}.} Therefore, it may be interesting to consider new bootstrap-based inference procedures that are robust to many IVs, many controls, and heteroskedastic errors simultaneously. 



\bibliographystyle{ecta}
{\small \bibliography{Biblio_boot_few_clusters}}

\newpage

\appendix

\begin{center}
    
{\Large{Supplementary Appendix to ``Wild Bootstrap Inference for Linear Regressions with Many Covariates"} }

\end{center}

\setcounter{table}{0}
\renewcommand{\thetable}{A\arabic{table}}

Section \ref{sec:diss_assumption} gives further discussion on the assumptions in Section \ref{sec:theory}. 
Section \ref{sec:pf_boot-stud-t-power} contains the proof of Theorem \ref{theom: 1}. 
Section \ref{sec:further simu} presents the simulation results for the one-way fixed effect model. 

The following notations are used for the bootstrap asymptotics: for any bootstrap statistic $T^*$, we write $T^* = o_{p^*}(1)$ in probability if for any
$\delta >0$, $\epsilon>0$, $\lim_{n \rightarrow \infty} P[P^*[|T^*|>\delta]>\epsilon]=0$, i.e., $P^*[|T^*|>\delta] = o_p(1)$. 
To be concise, we use the short version $T^* = o_{p^*}(1)$ to say that $T^* = o_{p^*}(1)$ in probability.  
Also, we write $T^* \rightarrow^{d^*} T$ in probability if, conditional on the sample, $T^*$ weakly converges to $T$ under $P^*$, for all samples contained in a set with probability approaching one. 
Specifically, we write $T^* \rightarrow^{d^*} T$ in probability if and only if $E^*[ f(T^*) ] \rightarrow E[ f(T) ]$ in probability for any bounded and uniformly continuous function $f$, 
where $E^*$ denotes the expectation under the probability measure induced by the bootstrap.

\section{Discussion on Assumptions}\label{sec:diss_assumption}

Assumptions \ref{ass: 1}-\ref{ass: 3} follow closely Assumptions 1-3 in \cite{cattaneo2018inference} and \cite{jochmans2022heteroscedasticity},  under which they show the asymptotic normality of the OLS estimator in the high-dimensional case with $q_n/n \nrightarrow 0$ as $n \rightarrow \infty$. 

Specifically, Assumption \ref{ass: 1} covers not only standard randomly sampled data but also repeated-measurement data 
such as short panel data.  $\{N_1,  ...,  N_{G_n}\}$ represents a partition of $\{1, ...,  n\}$ into $G_n$ strata,  which are independent from each other but dependence between observations within the strata is allowed. 
However,  this assumption does not allow for correlation in the error terms across units,  and therefore excludes clustered,  spatial or time series dependence in the sample.   

Assumption \ref{ass: 2} contains standard rank and moment conditions.  In addition,  it allows $q_n$ to grow at the same rate as the sample size.  

Furthermore, the settings in \cite{cattaneo2018inference} and \cite{jochmans2022heteroscedasticity} allow for the cases where the linear regression model in (\ref{eq: model}) is a linear-in-parameters mean-square approximation to the conditional expectation 
$\mu_i = E[y_i | X_n,  \mathcal{W}_n]$.  
Assumption \ref{ass: 3} provides regularity conditions on how fast such an approximation should improve.  
We refer interested readers to Section 3.2.3 of \cite{cattaneo2018inference} and Section 2.1 of \cite{jochmans2022heteroscedasticity} for more detailed discussions of this assumption.  

Assumption \ref{ass: 4} follows closely Assumption 4 of \cite{jochmans2022heteroscedasticity},  which is required for the consistency of his variance estimator. 
In particular,  the condition that $\max_i \mu_i^2 = o_p(n)$ is used to control the variance of $\sum_{i=1}^n \hat{v}_i^2 (y_i \acute{u}_i)$,  as the variance of $y_i \acute{u}_i$ depends on $\mu_i^2$. 
Interested readers are referred to Section 2.2 and Supplementary Appendix A.2 of \cite{jochmans2022heteroscedasticity} for more discussions and related sufficient conditions.   

Assumption \ref{ass: 5} contains the conditions for the modified wild bootstrap procedure.   
Specifically,  it requires the random weights of the wild bootstrap to be independent from the sample,  
to have their mean equal to zero,  their variance equal to one,  and a finite fourth moment. 
Similar to Assumption \ref{ass: 4},  we need the last condition in Assumption \ref{ass: 5} to control the (conditional) variance of $\sum_{i=1}^n \hat{v}_i^2 (y^*_i \acute{u}^*_i)$.

\section{Proof of Theorem \ref{theom: 1}}
\label{sec:pf_boot-stud-t-power}

We first show that under the null, $n^{-1} \sum_{i=1}^n \hat{v}_i^2 (y_i^*\acute{u}^*_i - \sigma_i^{*2})
= o_{p^*}(1)$. To do so, we follow the steps of the proof for Theorem 1 in \cite{jochmans2022heteroscedasticity}.
Notice that 
\begin{align}\label{eq: proof_1}
n^{-1} \sum_{i=1}^n \hat{v}_i^2 (y_i^*\acute{u}^*_i - \sigma_i^{*2})
= n^{-1} \sum_{i=1}^n \hat{v}_i^2 (u_i^{*2} - \sigma_i^{*2}) + n^{-1} \sum_{i=1}^n \hat{v}_i^2 (y_i^*\acute{u}^*_i - u_i^{2*}),
\end{align}
where $\sigma_i^{*2} = E^*[u_i^{*2}] = a_n^2(\beta_0) \tilde{u}^2_i(\beta_0)$. 

For the first term on the right-hand side of (\ref{eq: proof_1}),  we have 
$E^*\left[ n^{-1} \sum_{i=1}^n \hat{v}_i^2 (u_i^{*2} - \sigma_i^{*2}) \right] =0.$
Furthermore,  for the variance (conditional on the data)
\begin{align*}
& E^*\left[ \left(n^{-1} \sum_{i=1}^n \hat{v}_i^2 (u_i^{*2} - \sigma_i^{*2}) \right)^2 \right] \notag \\
&= n^{-2} \sum_{i=1}^n \hat{v}_i^4 \left( E^*[u_i^{*4}] - \sigma_i^{*4} \right)  
= n^{-2} \sum_{i=1}^n \hat{v}_i^4 \left( a_n^4(\beta_0) E^*[\omega_i^{*4}] \tilde{u}^4_i(\beta_0) - 
(a_n^2(\beta_0) \tilde{u}^2_i(\beta_0) ) \right) 
= o_p(1),
\end{align*}
which follows from $a^2_n(\beta_0) = O_p(1)$,  $\tilde{u}^2_i(\beta_0)	 = O_p(1)$, 
$(\max_i |\hat{v}_i|/\sqrt{n})^2 = o_p(1)$,  $n^{-1}\sum_{i=1}^n\hat{v}_i^2 = O_p(1)$, 
and Assumption \ref{ass: 5}.
In particular, $n^{-1}\sum_{i=1}^n \hat{v}_i^2 \leq n^{-1}\sum_{i=1}^n v_i^2 
\leq 2(n^{-1} \sum_{i=1}^n Q_i^2) + 2(n^{-1}\sum_{i=1}^n V_i^2)
=O_p(\chi_n) + O_p(1) = O_p(1)$.

Therefore,  we have 
\begin{align}\label{eq: proof_3}
n^{-1} \sum_{i=1}^n \hat{v}_i^2 (u_i^{*2} - \sigma_i^{*2}) =o_{p^*}(1). 
\end{align}
	
Then, for the second term on the right-hand side of (\ref{eq: proof_1}),  we use the following decomposition:
\begin{align}\label{eq: proof_2}
& n^{-1} \sum_{i=1}^n \hat{v}_i^2 (y_i^*\acute{u}^*_i - u_i^{2*}) \notag \\
&= n^{-1} \sum_{i=1}^n \sum_{j \neq i} \hat{v}_i^2 u_i^* (A_n)_{ij} u_j^* 
 + n^{-1} \sum_{i=1}^n \hat{v}_i^2 ((A_n)_{ii} - 1) u_i^{*2} + n^{-1} \sum_{i=1}^n \sum_{j=1}^n \hat{v}_i^2 \hat{\mu}_i(\beta_0) (A_n)_{ij} u_j^*,
\end{align}
which follows from 
$(A_n)_{ij} = (H_n)_{ij} / (M_n)_{ii}$,  $(H_n)_{ij} = (M_n)_{ij} - (n^{-1} \sum_{k=1}^n \hat{v}_k^2)^{-1} (n^{-1} \hat{v}_i \hat{v}_j )$, 
and $\acute{u}_i^* = \sum_{j=1}^n ((H_n)_{ij}/(M_n)_{ii}) u_j^* = \sum_{j=1}^n (A_n)_{ij} u_j^*$. 
We aim to show that all the three terms on the right-hand side of (\ref{eq: proof_2}) are of $o_{p^*}(1)$. 

For the first right-hand side term, we notice that 
$E^*\left[ n^{-1}\sum_{i=1}^n\sum_{j \neq i} \hat{v}_i^2 u_i^* (A_n)_{ij} u_j^* \right] = 0$
and 
\begin{align*}
& E^*\left[ \left( n^{-1}\sum_{i=1}^n\sum_{j \neq i} \hat{v}_i^2 u_i^* (A_n)_{ij} u_j^*\right)^2 \right] \notag \\
& = n^{-2} \sum_{i_1=1}^n \sum_{i_2=1}^n \sum_{j_1 \neq i_1} \sum_{j_2 \neq i_2} 
\hat{v}_{i_1}^2 (A_n)_{i_1, j_1} E^*[ u_{i_1}^* u_{j_1}^* u_{i_2}^* u_{j_2}^*] (A_n)_{i_2, j_2} \hat{v}^2_{i_2} \notag \\
& \leq \left( \max_i \sigma_i^{*2} \right)^2 n^{-2} \sum_{i=1}^n \hat{v}_i^4 \sum_{j \neq i} (A_n)_{ij}^2  
   + \left( \max_i \sigma_i^{*2} \right)^2 n^{-2} \sum_{i=1}^n \sum_{j \neq i} \hat{v}_i^2 \hat{v}_j^2 (A_n)_{ij} 
   (A_n)_{ji} = o_p(1),
\end{align*}
which follows from $E^*[ u_{i_1}^* u_{j_1}^* u_{i_2}^* u_{j_2}^*] = 0$ except for the cases where 
(1) $i_1=i_2$ and $j_1=j_2$ , or (2) $i_1=j_2$ and $i_2=j_1$, 
$\sum_{j \neq i} (A_n)_{ij}^2 \leq \sum_{j=1}^n (A_n)^2_{ij} = (H_n)_{ii} (M_n)_{ii}^{-2} 
\leq (M_n)_{ii}^{-2}$,  
$\sum_{j \neq i} (A_n)_{ij} (A_n)_{ji} \leq \sum_{j=1}^n (A_n)_{ij} (A_n)_{ji} 
\leq (\min_i (M_n)_{ii} )^{-2}$, 
$(\min_i(M_n)_{ii})^{-1}=O_p(1)$, 
$\max_i \sigma_i^{*2}=O_p(1)$,  
$(\max_i |\hat{v}_i |/\sqrt{n})^2 = o_p(1)$,  
and $n^{-1}\sum_{i=1}^n\hat{v}_i^2=O_p(1)$.

Using similar arguments,  we find the second right-hand side term in (\ref{eq: proof_2}) has (conditional) mean
\begin{align*}
E^* \left[ n^{-1} \sum_{i=1}^n \hat{v}_i^2 ((A_n)_{ii} -1) u_i^{*2} \right]
= \left( n^{-1} \sum_{i=1}^n \hat{v}_i^2 \right)^{-1} n^{-2} \sum_{i=1}^n \hat{v}_i^4 \sigma_i^{*2} (M_n)_{ii}^{-1} = o_p(1), 
\end{align*}
by using $n^{-1}\sum_{i=1}^n\hat{v}_i^2=O_p(1)$, $(\max_i |\hat{v}_i |/\sqrt{n})^2 = o_p(1)$,  
$\max_i \sigma_i^{*2}=O_p(1)$, and $(\min_i(M_n)_{ii})^{-1}=O_p(1)$. 
Similarly, we obtain that it has (conditional) variance
\begin{align*}
& E^* \left[ \left( n^{-1} \sum_{i=1}^n \hat{v}_i^2 ((A_n)_{ii} -1) u_i^{*2} \right)^2 \right] \notag \\
& = \left(n^{-1} \sum_{i=1}^n \hat{v}_i^2 \right)^{-2}
\left(n^{-4} \sum_{i=1}^n \sum_{j=1}^n \hat{v}_i^4 \hat{v}_j^4 (M_n)_{ii}^{-1} (M_n)_{jj}^{-1} E^*[u_i^{*2}u_j^{*2}] \right) = o_p(1), 
\end{align*}
by using the Cauchy-Schwarz inequality and Assumption \ref{ass: 5}. 

The third right-hand side term in (\ref{eq: proof_2}) has (conditional) mean
$E^*\left[ n^{-1} \sum_{i=1}^n \sum_{j=1}^n \hat{v}_i^2 \hat{\mu}_i(\beta_0) (A_n)_{ij} u_j^* \right] =0,$
and (conditional) variance
\begin{align*}
& E^*\left[ \left( n^{-1} \sum_{i=1}^n \sum_{j=1}^n \hat{v}_i^2 \hat{\mu}_i(\beta_0) (A_n)_{ij} u_j^* \right)^2 \right]  = \sum_{j=1}^n \sigma_j^{*2} \left( n^{-1} \sum_{i=1}^n \hat{v}_i^2 \hat{\mu}_i(\beta_0) (A_n)_{ij} \right)^2  \notag \\
& \leq \left( \max_{i} \sigma_i^{*2} \right) \left( \min_i (M_n)_{ii} \right)^{-2} 
\left( \max_{i} |\hat{\mu}_i(\beta_0) |/\sqrt{n} \right)^2 \left(n^{-1}\sum_{i=1}^n \hat{v}_i^4\right) = o_p(1),
\end{align*}
under the null hypothesis, by using $\max_i \sigma_i^{*2}=O_p(1)$, $(\min_i(M_n)_{ii})^{-1}=O_p(1)$, 
$\max_{i} |\hat{\mu}_i(\beta_0) |/\sqrt{n} = o_p(1)$, 
and $n^{-1} \sum_{i=1}^n \hat{v}_i^4 = n^{-1} \sum_{i=1}^n (\tilde{Q}_i + \tilde{v}_i)^4 
\leq 4 (n^{-1}\sum_{i=1}^n \tilde{Q}_i^4) + 4 (n^{-1}\sum_{i=1}^n \tilde{v}_i^4) = O_p(1)$,  
which follows from Assumptions \ref{ass: 1}, \ref{ass: 2}, and \ref{ass: 4}.

Therefore,  we obtain that all the three right-hand side terms in (\ref{eq: proof_2}) are of $o_{p^*}(1)$ and thus
$n^{-1}\sum_{i=1}^n\hat{v}_i^2(y_i^*\acute{u}_i^* - u_i^{*2}) = o_{p^*}(1)$. 

Combining with (\ref{eq: proof_3}),  we obtain that under the null,
$n^{-1} \sum_{i=1}^n \hat{v}_i^2 (y_i^*\acute{u}^*_i - \sigma_i^{*2}) = o_{p^*}(1)$, which, together with the definition of $\sigma_i^{*2}$, 
further implies that 
\begin{align}\label{eq: proof_4}
n^{-1} \sum_{i=1}^n \hat{v}_i^2 y_i^*\acute{u}^*_i - n^{-1} \acute{\Sigma}_n(\beta_0) = o_{p^*}(1).
\end{align}

Now, let $\bar{t}_n^* = (\hat{\beta}_n^* - \beta_0)/\sqrt{\acute{\Omega}_n(\beta_0)}$,  
where $\acute{\Omega}_n(\beta_0) = (\sum_{i=1}^n \hat{v}_i^2)^{-1}  \acute{\Sigma}_n(\beta_0) (\sum_{i=1}^n \hat{v}_i^2)^{-1}$. Then,  under the null hypothesis,  by Lemma 2.9.5 of \cite{van1996weak} and the bootstrap data generating process in Section 3,  we have 
\begin{align*}
\bar{t}^*_n = \left( n^{-1} \acute{\Omega}_n(\beta_0) \right)^{-1/2} \left( \sum_{i=1}^n \hat{v}_i^2 \right)^{-1} \frac{1}{\sqrt{n}}\sum_{i=1}^n \hat{v}_i u_i^*
\rightarrow^{d^*} N(0,1),  
\end{align*}
in probability.  
Furthermore,  by (\ref{eq: proof_4}) and Slutsky's theorem,  we have
\begin{align*}
t_n^* = \left( n^{-1} \acute{\Omega}_n^*\right)^{-1/2}  \left( n^{-1} \acute{\Omega}_n(\beta_0) \right)^{1/2} \bar{t}_n^* \rightarrow^{d^*} N(0,1),  
\end{align*}
in probability.  
Finally,  the conclusion follows by Polya's theorem.  $\blacksquare$

\section{Simulation Results for the Fixed Effect Model}\label{sec:further simu}

The second simulation model considered is a one-way fixed effect model for panel data, similar to the one considered in \cite{jochmans2022heteroscedasticity}. 
For double-indexed data $(y_{(g,m)}, x_{(g,m)})$,  it can be written as 
\begin{align*}
y_{(g,m)} = x_{(g,m)} \beta + \alpha_g + \epsilon_{(g,m)}, \;\; g=1, ..., G, \;\; m=1, ..., M, 
\end{align*}
where $\alpha_g$ is a group-specific intercept.
For this model, the fixed effect estimator equals the OLS estimator of $y_{(g,m)}$ on $x_{(g,m)}$ and $G$ group dummy variables. 
Following \cite{jochmans2022heteroscedasticity}, we let $x_{(g,m)} \sim i.i.d. N(0,1)$, $\epsilon_{(g,m)} \sim i.i.d.N(0,1)$, $\beta=2$, 
and $\alpha_g=0$ for all $g$. 
The total sample size $n = G \times M$ is set at $100$, and we set the number of groups $G \in \{5, 10, 20, 25, 50\}$. 
The simulation results are reported in Table \ref{tab:A1}. 
We observe that the tests based on $HC0$, $HCK$, and $HCA$ all tend to over-reject, especially in the case 
with $G=50$, while the two wild bootstrap procedures also exhibit better finite sample performance. 

\begin{table}[H]
\begin{center}
\begin{tabular}{llllll}
G & 5 & 10 & 20 & 25 & 50 \\ 
\hline
\hline
HC0    & 0.069 & 0.081 & 0.093 & 0.109 & 0.191 \\
HCK    & 0.064 & 0.067 & 0.066 & 0.071 & 0.191 \\
HCA    & 0.107 & 0.102 & 0.105 & 0.110 & 0.123 \\
Wild-G & 0.046 & 0.045 & 0.043 & 0.042 & 0.042 \\
Wild-R & 0.038 & 0.037 & 0.037 & 0.035 & 0.036 \\
\hline
\end{tabular}
\caption{Null rejection frequencies for Panel Data}
\label{tab:A1}
\end{center}
\end{table}

\end{document}